\documentclass[traditabstract]{aa}

\usepackage{amsfonts}
\usepackage{amsmath}
\usepackage{amssymb}
\usepackage{pdflscape}
\usepackage{rotating}
\usepackage{natbib}
\usepackage{txfonts}
\usepackage{longtable}
\usepackage[small,bf]{caption}
\usepackage[small,bf]{subfigure}
\usepackage{multirow}
\usepackage{xspace}
\usepackage{graphicx}
\usepackage{epstopdf}
\usepackage[pdftitle={3c120}, colorlinks=true, citecolor=blue, linkcolor=blue, urlcolor=blue]{hyperref}

\newcommand\simlt{\lower.5ex\hbox{$\; \buildrel < \over \sim \;$}}
\newcommand{\fermi}{{\it Fermi}-LAT\xspace}
\newcommand{\gray}{$\gamma$-ray\xspace}
\newcommand{\grays}{$\gamma$-rays\xspace}

\newcommand{\rg}{3C 120\xspace}
\DeclareUnicodeCharacter{B0}{\textdegree}

\begin{document}
\title{On the gamma-ray emission from \rg}
\titlerunning{\grays from \rg}

\author{N. Sahakyan\inst{1,2}
\and D. Zargaryan\inst{2}
\and V. Baghmanyan\inst{2}
}
 \institute{ICRANet, Piazza della Repubblica 10, I-65122 Pescara, Italy
 \and ICRANet-Yerevan, Marshall Baghramian Avenue 24a, Yerevan 0019, Republic of Armenia }

\abstract{We report the analysis of Fermi Large Area Telescope data from five years of observations of the broad line radio galaxy \rg. The accumulation of larger data set results in the detection of high-energy \grays up to 10 GeV, with a detection significance of about $8.7\sigma$. A power-law spectrum with  a photon index of $2.72\pm0.1$ and integrated flux of $F_{\gamma}=(2.35\pm0.5)\times10^{-8}\:\mathrm{photon\:cm}^{-2}s^{-1}$ above 100 MeV well describe the data averaged over five year observations. The  variability analysis of the light curve with 180-, and 365- day bins reveals flux increase (nearly twice from its average level) during the last year of observation. This variability on month timescales indicates the compactness of the emitting region. The \gray spectrum can be described as synchrotron self-Compton (SSC) emission from the electron population producing the radio-to-X-ray emission in the jet. The required electron energy density exceeds the one of magnetic field only by a factor of 2 meaning no significant deviation from equipartition.}

\maketitle

\section{Introduction}
The \gray detection by \fermi from non-blazar active galactic nuclei (e.g. Cen A \citep{abdo2010a}, M87 \citep{abdom87}, Ngc 1275 \citep{abdo2010b}) shows that these are different and potentially very interesting classes of \gray emitters. This provides an alternative approach to study high energy emission processes compared with blazars where the emission is strongly Doppler boosted. \\
At the red shift $z=0.033$, \rg is a nearby Seyfert 1 radio galaxy which is active and powerful emitter of radiation at all the observed wavebands. Having a bright continuum and broad optical emission lines \rg is usually classified as a Broad Line Radio Galaxy (BLRG). Hosting a black hole with a mass $5.5\times10^{7}M_\odot$, well constrained from the reverberation mapping \citep{peterson04}, it has a radio morphology more similar to the Fanaroff-Riley class I radio sources \citep{fanaroff74}. The source has a powerful one sided radio jet extending from a sub-$pc$ up to 100 $kpc$ scales \citep{walker}. Observations with Very Long Baseline Array at frequencies (22, 43, and 86 GHz) reveal a very rich inner jet structure containing several superluminal components with apparent speed up to 4-6 $c$ \citep{homan01,gomes98,gomes99} that can be investigated with better resolution than most other extragalactic superluminal sources because of the relatively low redshift. The jet inclination angle to the line of sight is limited to be $14^{\circ}$ by the apparent motion \citep{eracleus}. Recently, using X-ray and radio observations, \citep{mars02} found that dips in the X-ray emission are followed by ejections of bright superluminal knots in the radio jet which clearly establishes an accretion-disk-jet connection.\\
In X-rays \rg is a bright ($\approx5\times10^{-11}\:\mathrm{erg\:cm^{-2}s^{-1}}$ at 2-10 kev) and variable source on time scales from days to months \citep{halpern}. The {\it ASCA} observation shows a broad iron line $K_\alpha$ which can be fitted by Gaussian with $\sigma=0.8$ keV and equivalent width of 400 eV. The knots in the jet of \rg observed in radio and optical bands later have also been detected in the X-ray band with the {\it ROSAT} and {\it chandra} \citep{hariss04} that indicates the existence of high energy nonthermal particles in these knots. The origin of X-ray emission is highly debated especially when the extrapolation of synchrotron emission fails to account this emission. In these cases it can be explained by the inverse-Compton scattering of Cosmic Microwave Background (CMB) photons or by proton synchrotron emission, if so this component can be extended up to MeV/TeV range \citep{zhang,ahar2002}.\\      
At High Energies (HE; $>100$ MeV) the source has not been detected with the EGRET on board the Compton Gamma Ray Observatory even though with several pointing observations \citep{lin93}. The $2\sigma$ upper limit on the source flux above 100 MeV is set to $ 9\times 10^{-8}$\,cm$^{-2}$\,s$^{-1}$. Afterwards, the source is detected with the Fermi Large Are Telescope (\fermi) using 15 months of all sky exposures \citep{abdo10c}. The averaged HE spectrum between 100 MeV and 1 GeV can be described by the power-law with photon index $\Gamma=2.71\pm0.35$ and an integral flux $F(E>100\mathrm{\,MeV}) = (2.9 \pm 1.7_{\mathrm{stat}}) \times 10^{-8}$\,cm$^{-2}$\,s$^{-1}$ with the detection significance of $5.6\sigma$. However the source is not included in the \fermi second source catalogue \citep{abdo11} (2FGL) since the averaged signal appeared to be below the required $5\sigma$ threshold. This might be an evidence of the long term variability of the flux, since for a steady \gray signal the accumulation of longer data set (24- versus 15- months) should result an increase of the detection significance $\approx\sqrt{24/15}\times5.6\sigma$ what was not detected. \\
Also the temporal variations of the \gray flux (above 100 MeV) have been investigated in short time scales \citep{abdo10c,katak}. The light curve binned in the three-month-long periods shows only few episodes (one for 15- month \citep{abdo10c} and 2 for 24- month \citep{katak} data sets) when the flux increased on the level more than  $3\sigma$ and the rest of the time the source was undetectable by \fermi. This has been interpreted as a GeV flux variation on 90- day scales. However considering only 2 periods out of 8 has been detected and taking into account trials, post-trial significance probably is even lower than $3\sigma$. Therefore the conclusions in this regards were inconclusive considering the limited/poor statistics of the detected signal.\\
The peculiar structure of \rg implies different sites (sources) as the origin of the detected GeV \grays. Clearly, analogous with blazars, the non-thermal beamed radiation can be produced in the innermost part of the jet on the scales less than kilo-parsec (but not strongly boosted because of the larger jet inclination angle compared to blazars). In fact, a power-full jet observed at an inclination angle $\theta\leq14^{\circ}$, producing radio to x-ray flux (with the luminosity $\sim10^{45}\:\mathrm{erg\:s^{-1}}$) via synchrotron emission can produce Doppler boosted \gray flux via inverse Compton scattering. Such an emission would appear as variable on week time scales or shorter with the luminosity $L_{\gamma}\sim\: U_{rad}/U_B\:L_{syn}$ which in principle can be detected by \fermi (depending on the magnetic field and emitting region size). On the other hand, the non variable \gray emission from the extended structures (e.g. extended lobes, large scale jet and knots) can extend up to the GeV energies and give (at least at some level) a contribution to the total observed \gray flux. For example, a \gray emission from the extended lobes of nearby Centaurus A radio galaxy contributes greater than one-half of the total source emission \citep{abdolobes,yang}. Moreover, steady \gray emission from \rg on the level of \fermi sensitivity has already been predicted from the jet knots within both proton synchrotron \citep{ahar2002} and beamed CMB inverse Compton scenarios \citep{zhang}. Therefore the compact and extended structures both remains possible sites for production of the observed \grays.\\
In general the absence of statistical significant indication of variability introduces uncertainties to distinguish between different emission mechanisms. In particular for \rg, giving presence of many prominent sites for \gray production, the possible variability (or non variability) is crucial for understanding the origin of HE emission. Moreover, the spectrum extending only up to 1 GeV does not provide any possibility to differ the mechanisms using the predicted different spectral shapes. Now the larger data set allows to study the spectrum with better statistics above 1 GeV as well as a detailed investigation of the variability. This motivated us to have a new look on the \gray emission based on the five years of \fermi data.\\
The paper is structured as follows. The results of spectral analysis are presented in Sect. \ref{sec2} whereas the temporal analysis in Sect. \ref{sec3}. Implications of different emission mechanisms are discussed in the Sect. \ref{sec4} and conclusions are presented in the Sect. \ref{sec5}.
\section{Fermi-LAT Data Analysis}\label{sec2}
\subsection{Data Extraction}\label{sec2.1}
\fermi on board the Fermi satellite is a pair-conversion telescope, operating since August 4, 2008 and is designed to detect HE $\gamma$-rays in the energy range 20 MeV - 300 GeV \citep{atwood09}. It constantly scans the entire sky every three hours and by default is always in the survey mode. Details about the LAT instrument can be found in \citep{atwood09}.\\
For the present analysis we use publicly available \fermi $\sim5.3$~yr data from 4th August 2008 to 4th December 2013 (MET 239557417--407808003). We use the Pass 7 data and analyze them using the Fermi Science Tools  v9r33p0 software package.  Events with zenith angle $<100^{\circ}$ and with energy between 100 MeV and 100 GeV were selected. Only the data when the rocking angle of the satellite that was $<52^{\circ}$, are used to reduce the contamination from the Earth limb $\gamma$-rays, which are produced by cosmic rays interacting with the atmosphere. We download photons from a $10^{\circ}$ region centered on VLBI radio position of \rg (RA,dec)= (68.296, 5.354) and work with a $14^{\circ}\times14^{\circ}$ square region of interest (ROI). Photons are binned with {\it gtbin} tool with a stereographic projection into pixels of $0.1^{\circ}\times0.1^{\circ}$ and into 30 equal logarithmically-spaced energy bins. Then with the help of {\it gtlike} tool a standard binned maximum likelihood analysis is performed. The fitting model includes diffuse emission components and $\gamma$-ray sources within ROI (the model file is created based on the \fermi second catalog \citep{abdo11}) and since \rg is not included in 2FGL we added a point like source on the known location of \rg (RA,dec)= (68.296, 5.354) \citep{ma}. The Galactic background component is  modeled using the LAT standard diffuse background model {\it gll\_ iem \_ v05\_ rev1}  and {\it iso\_source\_v05} for the isotropic $\gamma$-ray background. The normalization of background models as well as fluxes and spectral indices of sources within $10^{\circ}$ are left as free parameters in the analysis.
\subsection{Spectral Analysis}\label{sec2.2}
\begin{figure}
   \centering
  \includegraphics[width=0.5 \textwidth]{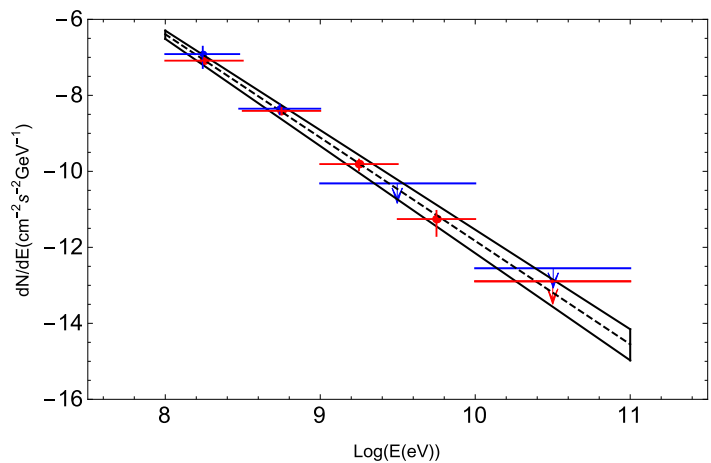}
   \caption{ The averaged differential spectrum of \rg (above 100 MeV) red points - this work -as compared with the one based on the initial 15 month data set \citep{abdo10c}. The dashed black line shows the power-law function determined from the {\it gtlike}.}
    \label{fg1}
\end{figure}
We assume that the $\gamma$-ray emission from \rg is described by the power-law and normalization and power-law index are considered as free parameters, then the binned likelihood analysis is performed. From a binned {\it gtlike} analysis, the best-fit power-law parameters for \rg are
\begin{equation}
\left(\frac{dN}{dE}\right)_{P}=(4.06\pm1.0)\times10^{-10}\left(\frac{E}{100\;\mathrm{MeV}}\right)^{-2.72\pm0.1}\,.
\label{pl}
\end{equation}
This corresponds to an integral flux of
\begin{equation}
F_{\gamma}=(2.35\pm0.5)\times10^{-8}\:\mathrm{photon\:cm}^{-2}s^{-1},
\label{pl1}
\end{equation}
with only statistical errors taken into account. The test statistic (defined as TS = 2(log $L$ - log $L_0$), where $L$ and $L_0$ are the likelihoods when the source is included or not) is $TS$ = 76.3 above 100 MeV, corresponding to a $\approx 8.7\: \sigma$ detection significance. The results are consistent with the parameters found in \citep{abdo10c}, namely photon index $\Gamma=2.71\pm0.35$ and integral flux $(2.9\pm1.7)\times10^{-8}$ ph cm$^{-2}$s$^{-1}$ above $100$ MeV. The value $TS$ = 76.3 is above the threshold value TS=25 and \rg should be included in the upcoming \fermi source catalogs.\\
Figure~\ref{fg1} shows the spectrum of \rg obtained by separately running {\it gtlike} for 5 energy bands, where the dashed line shows the best-fit power-law function for the data given in Eq.~(\ref{pl}). For comparison results from the previous study of \rg \citep{abdo10c} are presented as blue data points. For the highest energy bin (10-100 GeV), it is shown an upper limit.\\
Since we used the exposure of almost 2.5 times longer than used in 2 FGL, this can result additional faint sources in the data which are not properly accounted in the model file. In order 
to check if any additional sources were present, the {\it gttsmap} Fermi tool is used with the best-fit model of 0.1- 100 GeV events to create a TS significance map of the $6^{\circ}\times6^{\circ}$ region. Although, no significant excess hot spots (TS$>$ 25) are presented. Therefore, the model file used in the analysis gives a well representation of the data.
Next, we obtained the source localization with {\it gtfindsrc}, yielding R.A. = 68.205, decl. = 5.38 with a 95\% confidence error circle radius of $r_{95} = 0.05$. These localizations are offset by $0.09^{\circ}$ from the VLBI radio position of \rg (R.A. = 68.296, decl. = 5.354) \citep{ma}.
\section{Temporal Variability}\label{sec3}
The variability of the observed \gray flux could provide important constraints on the emitting region(s). The time scale of the observed flux variation $\tau$ would limit the (intrinsic) size of the \gray production region to $R'/\delta\leq \frac{c \: \tau}{1+z}$ where $\delta$ is the Doppler factor and $z$ is the red shift. During the previous variability study, using the accumulation of 90- day \fermi all-sky-survey exposures, the source shows two time intervals when $TS>10$ whereas it was mostly being undetectable by \fermi. This could be  interpreted as a possible variability at GeV energies \citep{katak}. \\
\begin{figure}
   \centering
    \includegraphics[width= 0.5 \textwidth]{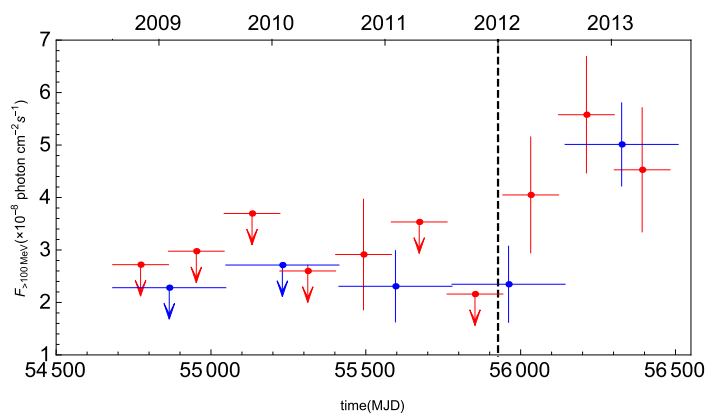}
    \caption{\gray light curve from August 4th 2008 to December 4th 2013. The bin size corresponds to 180- (red) and 365- days (blue). The galactic and extragalactic background emission is fixed to the best-fit parameters obtained for the overall time fit.}%
    \label{fg2}
\end{figure}
Now, more observational data set (increased photon statistics) can provide more details in this regards. Accordingly the total observational time (from August 4th, 2008 to December 4th, 2013) is divided into different timescales and light curves are generated using the unbinned likelihood analysis with {\it gtlike}.  In order to minimize uncertainties in the flux estimation, the photon indices  of all sources are fixed to the values obtained in 100 MeV-100 GeV energy range for the whole time period. Instead, the normalization of \rg and background point sources are treated as free parameters. Since no variability is expected from the underlying background diffuse emission, the normalization of both background components is fixed to the values which were obtained for the whole time period.\\
In Fig. \ref{fg2} is shown the \gray flux variation above 100 MeV for 180- and 365- days sampling with red and blue data points respectively. It is noticeable that up to $\approx55400$ MJD, \grays from \rg are below \fermi sensitivity (detection significance TS$<$ 10). Afterwards a cycle with faint \gray emission changes and then the produced flux is sufficient to be detected by \fermi. For example,  the averaged flux in a year sampling is more than twice higher than its average level (see Fig. \ref{fg2}) with highest test statistics corresponding to $TS=63.8$ (similarly for 180- day sampling  $TS=38.92$). Probably this is caused by the changes in emission states e.g. the source moves in a state which is characterized by more effective production of \grays resulting a flux increase. This flux increase gives proof of flux variability on month timescales.\\
In addition to reported month time scale variability, flux variation for shorter time periods e.g. a month or sub-month time scales has been performed over the time interval where the increase of flux was detected. Accordingly the unibinned likelihood analysis is performed using shorter time sampling (7 - and 15 - days) for the time period $>56000$ MJD. Interestingly, a \gray signal from the source is detected above threshold $TS>10$ using a sampling of only 7- days. Required condition was fulfilled only in three time intervals with the maximum detection significance $4.8\sigma$ reached in the last week of September 2013.  The corresponding flux is $(2.15\pm0.6)\times10^{-7}\mathrm{photon\:cm}^{-2}s^{-1}$ nearly an order of magnitude higher than average flux level presented in Eq. \ref{pl1}.  Because of limited statistics, however, no definite conclusion on shorter time scale variability can be drawn.
\section{Disscusion and Interpretation}\label{sec4}
The month time scale variability of \rg ($t_{var}\sim$ 6 month) denotes the compactness of the emitting region. Under any reasonable assumption for Doppler boosting, $\delta=3-5$, the emitting region cannot be larger than $R<\delta\:c\:t_{var}\sim10^{18}\:(\delta/4)\,\mathrm{cm}$. This immediately allows us to exclude jet knots as the main sites where the observed \grays are produced. Most likely the \grays are produced in a compact region of the jet e.g. the blob moving with relativistic velocities. Generally the  broad band spectrum of blazars (as well as from radio-galaxies which have jets oriented at systematically larger angles to our line of sight) are successfully described by the Synchrotron/Synchrotron self-Compton (SSC) \citep{ghise85,celot91,bloom96} model. In this modeling the low energy emission (radio through optical) is represented as a synchrotron emission from leptons in the homogeneous, randomly oriented magnetic field ($B$) while HE component (from x-ray to HE \gray) is an inverse Compton scattering of the same synchrotron photons. This kind of interpretation for \rg is the first choice considering the results of the modeling of the other \fermi observed radio galaxies \citep{abdo2010a,abdom87,abdo2010b}. Here we apply  SSC mechanism to model the overall SED of \rg particularly in the 0.1-100 GeV energy range. The multifrequency data (sub MeV/GeV energies) are from the simultaneous (quasi simultaneous) observations of the \rg \citep{giommi}.\\
We suppose that the emission is coming from a spherical region with the radius $R_b$ moving with Lorentz factor $\Gamma=(1-\beta)^{-1/2}$. The emission is boosted by $\delta=1/[\Gamma(1-\beta \cos(\theta))]$ where $\theta$ is the angle between the bulk velocity and the line of sight. The electron distribution follows $N(\gamma)\propto\gamma^{-\alpha}Exp[-\gamma/\gamma_c]$ with $\gamma>\gamma_{min}$ naturally expected from the shock acceleration theories and the electron energy density ($U_e$) scales with the one of the magnetic field ($U_B$). Then, several independent parameters used in the modeling can be constrained from the observations. In particular, the superluminal speed puts an upper limit to the jet`s inclination angle to be $14^{\circ}$ \citep{eracleus}. Thus, the flux would be modestly boosted  with the Doppler factor $\delta=4$ if the emitting region moves with the bulk Lorentz factor $\Gamma=8$ at the inclination angle of $12.5^{\circ}$. Next, monthly timescale variability ($180-$ days) implies that the emitting region is confined to a volume which radius is determined from the relation $R_b/\delta<\boldsymbol{4.6\times 10^{17}}$ cm. The Doppler factor of $\delta\sim(3-4)$ requires an emitting region with the size  $\sim10^{18}$ cm.\\
\begin{table}
\caption{SSC modeling parameters presented in Figure \ref{fg3}. The Doppler boosting is assumed to be $\delta=4$ and emitting region radius $R_b\sim\boldsymbol{10^{18}}$ cm.}
\centering
\begin{tabular}{c c c c c c c}
\hline\hline
  & \small{$B$(mG)} & \small{$\alpha$} & \small{$\gamma_{min}$} &\small{$\gamma_{c}$} &\small{ $U_e/U_B$} \\ [0.5ex]
\hline
\small{solid line} & 30 & 2.4 & 700 & $1.7\times10^{4}$ & 1.9\\
\small{dashed line}  & 25 & 2.8 & 800 & $9.8\times10^{3}$ & 16 \\
\small{dot dashed line} & 25& 2.0 & 2500 & $1.2\times10^{4}$ & 1.3  \\[1ex]
\hline
\end{tabular}
\label{tab2}
\end{table}
\begin{figure}
   \centering
    \includegraphics[width= 0.5\textwidth]{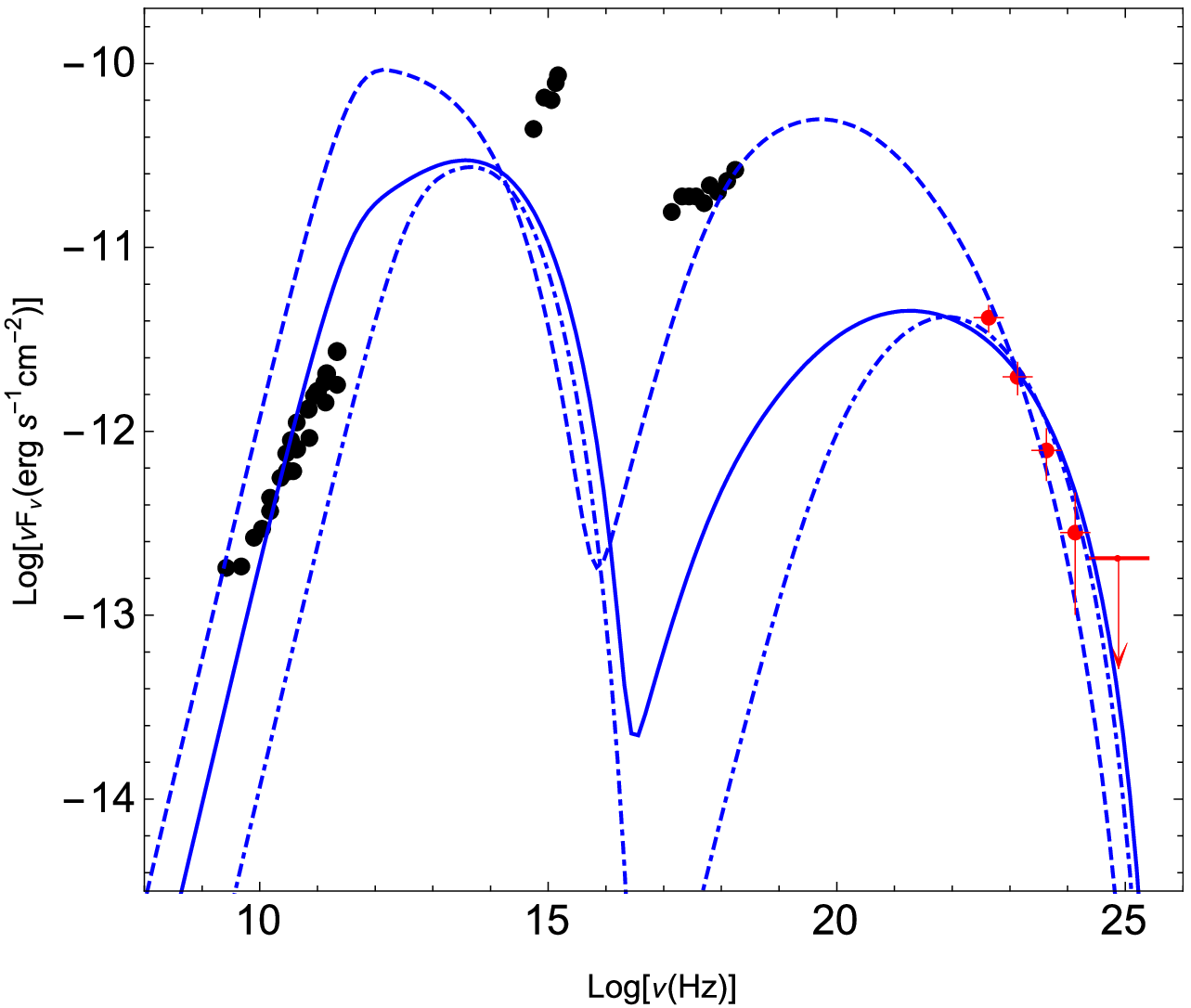}
    \caption{ The SED of \rg modeled with one-zone SSC component. Black points refers to the simultaneous (quasi-simultaneous) data from \citep{giommi} and red points from \fermi data analysis (this work). The solid, dashed and dot dashed lines corresponds to $\alpha=2.4, 2.8$ and $2$ respectively (see text for details). The SSC emission is calculated using a simulator developed by Andrea Tramacere \citep{code1,code2,code3} available at $http://www.isdc.unige.ch/sedtool/$.
  }%
    \label{fg3}
\end{figure}
We assume that HE emission has a pure SSC origin. As a first step, the radio data are included in the modeling which means that the same electron population is responsible for both synchrotron and inverse-Compton emissions. The best guess values of electron energy distribution which allows us to properly reproduce the low and HE data corresponds to $\alpha=2.8$, $\gamma_{min}=800$ and  $\gamma_{c}=9.8\times10^3$ (dashed line in Fig. \ref{fg3}). Other model parameters are presented in Table \ref{tab2}. The ratio of non- thermal electron and magnetic field energy densities is equal to $U_e/U_B\approx16$ (for magnetic field $B=25$ mG). In this case the jet power in the form of the magnetic field and electron kinetic energy, given by $L_{B}=\pi c R_b^2 \Gamma^2 U_{B}$ and $L_{e}=\pi c R_b^2 \Gamma^2 U_{e}$, respectively, are $L_{B}=1.49\times10^{44}\:\mathrm{erg\:s^{-1}}$ and  $L_{e}=2.42\times10^{45}\:\mathrm{erg\:s^{-1}}$. The total jet power $L_{jet}$, defined as $L_{jet}=L_B+L_e$, corresponds to $L_{jet}\approx2.57\times10^{45}\:\mathrm{erg\:s^{-1}}$ which is noticeable high. One can disfavor such by modeling noting that the necessary jet power is of the same order as Eddington accretion power $L_{Edd}=6.8\times10^{45}\mathrm{erg\:s^{-1}}$ for the $5.5\times10^7M_{\odot}$ black hole mass in \rg. But this is not a strong argument considering that some blazars might operate in the super-Eddington regime as follows from observations \citep{bonnoli}.\\
However, this is not the case, since the jet power can be relaxed interpretending the hard X-ray emission originates from the thermal Comptonization near the disk. Thus, the predicted flux by SSC component falls below the hard X-ray limit in Fig. \ref{fg3} (solid line). Indeed, a reasonable good modeling of both radio and HE data gives an electron distribution with the index $\alpha=2.4$ between $\gamma_{min}=700$ and $\gamma_{cut}=1.7\times10^{4}$. The jet energy carried out by particles (electrons) and magnetic field corresponds to $L_{j}=6.27\times10^{44}\mathrm{erg\:s^{-1}}$ which is still high but not dramatic. Moreover unlike the other case the electron non-thermal and magnetic field energy densities are close to equipartition $U_e/U_B\approx2$ (for $B=30$ mG). Even though the ratio $U_e/U_B\approx16$ can not be rejected recalling other blazars where the jet is massively out of equipartition, the later modeling has an advantage considering required total jet power. \\
In principle the radio to X-ray and \gray emissions can be produced in a different sites (blobs). Supposing the radio flux does not exceed the one presented in  \citep{giommi} the electron distribution with a typical power-law index $\alpha=2$ predicted from strong shock acceleration theories can reproduce HE \gray data  (dot dashed line in Fig. \ref{fg3}). The modeling requires relatively high low energy cut-off $\gamma_{min}=2500$ and energy equipartition between nonthermal electrons and magnetic field $U_e/U_B=1.3$ (for $B=25$ mG). Moreover, the total  jet kinetic power is $L_{jet}\approx3.4\times10^{44}\:\mathrm{erg\:s^{-1}}$ approximately twice less than in the previous modeling. From the point of view of the necessary lower energy this model has an advantage over the previous modelings. Nevertheless this is very sensitive to the choice of the $\gamma_{min}$ which can be constrained only with the simultaneous data. Although, the radio data presented in Fig. \ref{fg3} are not synchronous, they can be treated as an upper limit. Consequently the expected luminosity should not be higher than the above obtained value. \\
In Fig. \ref{fg3} SSC mechanism provides a good fit to all data except those in the optical/UV band $(10^{15}-10^{16})$ Hz. This UV excess is likely caused by direct thermal emission from the accretion disc. Indeed, a thermal component with a black body temperature $>15000$ K  and a luminosity $\geq2\times10^{44}\:\mathrm{erg\:s^{-1}}$ can explain detected UV flux. This lower limit to the temperature and luminosity corresponds to minimal UV flux reported in \citep{giommi} and presented in Fig. \ref{fg3} but hotter and luminous disc is expected to explain observed data. Thus, SSC radiation plus thermal component (contribution of the accretion disc) can satisfactorily reproduce the entire SED (including UV data). However detailed modeling of the thermal component goes beyond the scope of this paper.
\section{Conclusion}\label{sec5}
We report on the recent observations of \rg with \fermi. The source is detected up to 10 GeV with statistically significant $8.7\sigma$ significance as a result of the accumulation of the data from longer all sky exposure. The photon index corresponds to $\Gamma=2.7$ similar to the nearby FR1 class radio source Centaurus A  with the comparable black hole mass \citep{abdo2010a} and $F_{\gamma}=(2.35\pm0.5)\times10^{-8}\:\mathrm{photon\:cm}^{-2}s^{-1}$ photon flux above 100 MeV. Adapting the $d_L=139.3$ Mpc distance, this equals to $L_{\gamma}=2.1\times10^{43}\mathrm{erg\:s^{-1}}$ which lies in the typical isotropic \gray luminosity range of FR I sources detected  by \fermi \citep{abdo10c}.  
Albeit, the observed \gray flux is relatively faint compared with other \fermi detected radio galaxies ($\approx10^{-7}\mathrm{photon\:cm}^{-2}s^{-1}$), the isotropic \gray luminosity is quite impressive when compared with the Eddington luminosity $L_{Edd}=6.8\times10^{45}\mathrm{erg\:s^{-1}}$ .\\ 
We report also an interesting modification of the \gray flux in time. Initially the source described by the \gray flux mostly below than the \fermi sensitivity threshold appears to be frequently detected afterwards. The flux is almost twice more than its average level in the last year of the selected time period (from 2008 to late 2013, see Fig. \ref{fg2}). This increase of flux shows monthly scale variability of \rg indicating that the \grays are produced in sup-parsec regions. A common behavior of the light curves in any day sampling (a month or more): the source is mostly undetectable by \fermi prior to March 2012 then it turns to be mostly in \gray production duty cycles. Moreover the long lasting source activity which probably continues after December 2013 (nearly 2 years)  indicates a change in the \gray production state (from low to high) rather than flaring activity as seen in many blazars (generally in short time scales). In principle this change can have different physical origins. First, the change in the central engine, where possible jets obtain much of their energy from the infall of matter into a supermassive black hole, can at least can have some influence. The change in the jet power, hence higher intensity \gray, which is expected in the case when the additional matter is fueling the accretion disk. In theory the observations of the region closer to the black hole with sensitive X-ray instruments (e.g. {\it chandra}, {\it XMM-Newton}) can prove such a possibility. On the other hand, the environmental influence on the changes in \gray emission states can not be rejected  considering the large scale powerful jet up to 100 $kpc$ (e.g. target interacting with the jet). Any of the above mentioned possibilities would be supported by multiwavelength observations. This is beyond the scope of this paper and will be investigated in the future works.\\ 
One zone SSC model is used to fit broadband emission from \rg. Assuming \grays are produced in a compact region ($\sim10^{18}$ cm) this modeling gives an adequate fit to the SED with modest Doppler boosting $\delta\sim4$ and no significant bias from equipartition $U_e/U_B \approx2$. The necessary jet kinetic power is $\approx6\times10^{44}\;\mathrm{erg\:s^{-1}}$ which corresponds to 10 \% of Eddington power.

\begin{acknowledgements}
We warmly thank Felix Aharonian for detailed and constructive comments that improved the manuscript.
\end{acknowledgements}
\bibliographystyle{aa}
\bibliography{biblio}{}
\end{document}